\documentclass{article}

\usepackage[T1]{fontenc} 
\usepackage{hyperref}
\usepackage[backend=biber,style=numeric]{biblatex}
\usepackage[version=4]{mhchem}
\usepackage{float}
\usepackage{tabularx}
\usepackage{arxiv}
\usepackage{amsmath,amssymb}
\usepackage[utf8]{inputenc} 
\usepackage[T1]{fontenc}    
\usepackage{hyperref}       
\usepackage{url}            
\usepackage{booktabs}       
\usepackage{amsfonts}       
\usepackage{nicefrac}       
\usepackage{microtype}      
\usepackage{lipsum}
\usepackage{graphicx}
\graphicspath{ {./images/} }

\title{CarAT: Carbon Atom Tracing across Industrial Chemical Value Chains via Chemistry Language Models}

\author{
\textbf{Emma Pajak$^{1}$, David Walz$^{2}$, Olga Walz$^{2}$, Laura Marie Helleckes$^{1}$,} \\
\textbf{Klaus Hellgardt$^{1}$, Antonio del Rio Chanona$^{1*}$} \\
$^{1}$Department of Chemical Engineering, Imperial College London, London SW7 2AZ, United Kingdom \\
$^{2}$BASF SE, Ludwigshafen, Germany \\
\texttt{a.del-rio-chanona@imperial.ac.uk}
}

\begin{document}
\maketitle
\begin{abstract}
The chemical industry is increasingly prioritising sustainability, with a focus on reducing carbon footprints to achieve net zero. By 2026, the Together for Sustainability (TfS) consortium will require reporting of biogenic carbon content (BCC) in chemical products, posing a challenge as BCC depends on feedstocks, value chain configuration, and process-specific variables. While carbon-14 isotope analysis can measure BCC, it is impractical for continuous industrial monitoring. This work presents CarAT (Carbon Atom Tracker), an automated methodology for calculating BCC across industrial value chains, enabling dynamic and accurate sustainability reporting. The approach leverages existing Enterprise Resource Planning data in three stages: (1) preparing value chain data, (2) performing atom mapping in chemical reactions using chemistry language models, and (3) applying a linear program to calculate BCC given known inlet compositions. The methodology is validated on a 27-node industrial toluene diisocyanate value chain. Three scenarios are analysed: a base case with fossil feedstocks, a case incorporating a renewable feedstock, and a butanediol value chain with a recycle stream. Results are visualised with Sankey diagrams showing the flow of carbon attributes across the value chain. The key contribution is a scalable, automated method for real-time BCC calculation under changing industrial conditions. CarAT supports compliance with upcoming reporting mandates and advances carbon neutrality goals by enabling systematic fossil-to-biogenic substitution. Through transparent, auditable tracking of carbon sources in production networks, it empowers data-driven decisions to accelerate the transition to sustainable manufacturing.
\end{abstract}


\section{Introduction}
The global chemical industry faces mounting pressure to achieve net-zero emissions targets as part of broader decarbonization efforts across all industrial sectors. Chemical manufacturing, which accounts for approximately 5\% of global greenhouse gas emissions, plays a critical role in this transition due to its dual position as both a significant emitter and an enabler of sustainable solutions across multiple value chains. Approaches to advancing sustainability in chemical engineering are varied, with recent developments spanning life cycle-based optimization frameworks \cite{lechtenberg_pulpo_2024}, waste valorization through brewing innovations \cite{merkouri_integrated_2025}, and intensified separation processes \cite{tian_process_2022} among others \cite{garcia-serna_new_2007, baldea_transforming_2025}. This drive towards sustainability has catalyzed the development of comprehensive frameworks for measuring, reporting, and ultimately reducing the carbon footprint of chemical products throughout their life cycle.\\
Leading this transformation is the Together for Sustainability (TfS) consortium, comprising major chemical manufacturers including BASF, Bayer, Evonik Industries, Henkel, Lanxess, and Solvay \cite{noauthor_tfs_pcf_guidelines_2024_en_pages-lowpdf_nodate}. TfS aims to set global standards for sustainability reporting within the chemical industry, building upon principles established by the United Nations Global Compact, the Responsible Care Global Charter, and standards set by the International Labor Organization and the International Organization for Standardization. The consortium has introduced specific requirements for chemical companies to report their products' BCC by 2026, creating an urgent need for standardized assessment methodologies.

\subsection{Value Chains in the Chemical Industry}
Value chains, first defined by Porter as a ``\textit{set of activities that are performed to design, produce, market, deliver, and support its product}'' \cite{porter_competitive_1985}, are fundamental to the chemicals industry. In this context, a value chain refers specifically to the underlying network of raw materials, production facilities, and products—a system of value-adding chemical reactions and transformations that derives high-value products from feedstock \cite{holweg_defining_2014}. While supply chains focus on the physical movement and logistics of materials, value chains encompass the entire value-creation process, including the chemical transformations that occur at each stage.\\
An intrinsic property of chemical value chains is their interconnectedness, arising from synthesis pathways that draw on reactants from various sources, thereby linking the pathways of different products \cite{blackburn_operations_2015}. These chains can be vast and complex, with some synthesis pathways relying on a series of intricate chemical transformations to produce desired compounds. Recycle streams add further complexity by creating additional loops and interactions within the production process. Major chemical manufacturers such as BASF, Dow Chemical, Shell Chemicals, and Mitsubishi Chemical operate extensive global value chains comprising numerous interconnected pathways, encompassing hundreds of thousands of nodes and yielding thousands of diverse commercial products \cite{basf_ludwigshafen_2023}.

\subsection{Challenges in Molecular-Level Sustainability Analysis}

A fundamental challenge in achieving comprehensive sustainability assessment lies in the molecular-level opacity of existing business systems. Value chains are typically represented through Enterprise Resource Planning (ERP) systems, which integrate various business processes and functions into a unified system. While ERP data encompasses procurement, manufacturing, logistics, sales, and finance operations, it does not provide immediate or transparent access to the molecular-level details of a chemical value chain. Although these systems track the flow of materials and products, they do not inherently reveal the specific molecular species involved at each node.

This lack of molecular transparency poses significant challenges for sustainability assessment and reporting. Without visibility into the chemical transformations occurring at each production stage, it becomes difficult to:

\begin{itemize}
    \item Track the origin and fate of carbon atoms through complex reaction networks.
    \item Assess the potential for substituting fossil-derived inputs with biogenic alternatives.
    \item Calculate accurate sustainability metrics that require molecular-level understanding.
    \item Identify optimization opportunities for reducing environmental impact.
\end{itemize}

\subsection{Sustainability Reporting Requirements}
At the core of sustainability reporting for chemical manufacturers is the Product Carbon Footprint (PCF), a metric that quantifies the total greenhouse gas emissions associated with a product throughout its life cycle \cite{noauthor_product_nodate}. PCF sums up the total greenhouse gas emissions generated by a product over different stages of its life cycle, measured in CO$_2$ equivalents (CO$_2$e), a standardized unit that expresses the global warming potential of various greenhouse gases relative to carbon dioxide.

The TfS Guidelines provide a specialized framework for calculating PCFs for chemical products, ensuring adherence to internationally recognized standards for greenhouse gas accounting and environmental assessment. These guidelines align with Principle 7 of Green Chemistry, which advocates for the use of renewable feedstocks rather than depleting ones, by providing mechanisms to track and incentivize the transition from fossil to biogenic carbon sources throughout chemical value chains.

A critical upcoming requirement from the TfS Guidelines is the reporting of a product's Biogenic Carbon Content (BCC), starting in 2026 \cite{noauthor_product_nodate, noauthor_tfs_pcf_guidelines_2024_en_pages-lowpdf_nodate}. Biogenic carbon is defined by the World Business Council for Sustainable Development as ``\textit{carbon derived from living organisms or biological processes, but not fossilized materials or fossil sources}'' \cite{noauthor_pathfinder-framework-version-20pdf_nodate}. Typical sources include trees, plants, and soil, which absorb CO$_2$ as a natural part of their life cycle \cite{harris_chapter_2018}.

\begin{figure}
    \centering
    \includegraphics[width=0.8\linewidth]{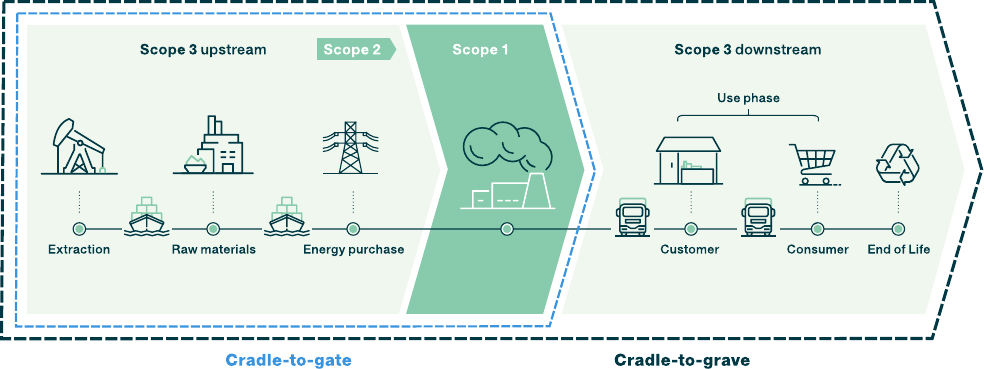}
    \caption{System boundary definition for Product Carbon Footprint (PCF) calculations showing the cradle-to-gate scope. The PCF includes Scope 1 (direct emissions from owned or controlled sources), Scope 2 (indirect emissions from purchased energy), and upstream Scope 3 emissions (from purchased goods and services). Downstream Scope 3 emissions from product use and end-of-life disposal are excluded from PCF calculations but can be estimated using BCC data. Figure adapted from TfS \cite{noauthor_tfs_pcf_guidelines_2024_en_pages-lowpdf_nodate}}
    \label{fig:pcf}
\end{figure}

The introduction of BCC reporting serves a crucial purpose: it supports the estimation of end-of-life emissions, which fall outside the cradle-to-gate scope of PCF calculations (see Figure \ref{fig:pcf}). The cradle-to-gate boundary encompasses all emissions from raw material extraction (Scope 3 upstream) through production processes (Scope 1 and 2) to the factory gate, but excludes downstream emissions from product use and disposal (Scope 3 downstream). For example, a product whose carbon is entirely biogenic would contribute no fossil CO$_2$ emissions through combustion or degradation, regardless of the end-of-life treatment. 

The ability to quickly calculate and recalculate BCC becomes increasingly critical as manufacturing landscapes evolve. As resilient supply chains expand the availability of biogenic raw materials \cite{paulo_approach_2023}, manufacturers need rapid BCC assessments for scenario analysis and credible end-of-life emission estimates. Similarly, as net-zero chemical pathways scale up ``\textit{carbon capture, low-carbon hydrogen, carbon storage, biomass utilization}'' \cite{gabrielli_net-zero_2023} and other technologies, manufacturers will require a BCC framework that can quickly update with changes in routes, feedstocks, and recycling to preserve transparent, auditable product-level attribution.

Calculating the BCC for products within complex value chains presents technical challenges. Currently, this calculation is only feasible when the chemical structure clearly differentiates biogenic carbon atoms from non-biogenic ones. For instance, in an ethoxylated fatty acid—typically synthesized from a fatty acid and ethylene oxide—the fatty acid component is generally derived from vegetable oil (biogenic), while the ethylene oxide may be fossil-derived. The biogenic content is then determined by the fraction of carbon originating from the fatty acid.

The complexity increases for products requiring multiple intermediates that may have their feedstocks replaced with biogenic or recycled materials. Accurate BCC calculation demands comprehensive knowledge of all upstream reactions, including value chain configuration, raw material composition, recycle streams, and process-specific nuances affecting product composition. As BCC depends on these dynamic variables, any upstream changes necessitate recalculation—a significant burden in global value chains where sustainable feedstocks, process setups, and efficiencies frequently change.

\subsection{The Need for Automated BCC Calculation}
Recent advances in automated PCF calculation methodologies have demonstrated the potential for transforming sustainability assessment in industrial settings. Notable examples include AllocNow \cite{3con2023sustainability}, CarbonMinds \cite{stellner2024carbon}, and Siemens' SiGreen \cite{hohlweck2023sigreen}, which have established automated frameworks for comprehensive carbon accounting across complex value chains. These tools demonstrate key advantages of automation: providing consistent methodologies, enabling certification and traceability, and facilitating scenario analysis for decision-making.

Chemical manufacturers face a critical challenge: they must obtain certification for their sustainability metrics, yet recalculating BCC each time upstream changes occur (such as feedstock modifications or process efficiency improvements) is impractical. An alternative approach involves developing a general computational methodology that decouples the dynamic variables of the value chain from the calculation process. By certifying the methodology itself, manufacturers automatically receive certification for all subsequent calculations performed using the approved procedure. BASF has successfully employed this strategy for PCF calculations through its SCOTT methodology \cite{noauthor_product_nodate}, creating a strong precedent for developing similar approaches for BCC.


\subsection{Aim and Scope}
This work aims to present a comprehensive framework for calculating the BCC of products within chemical value chains, enabling compliance with upcoming TfS sustainability reporting requirements. The framework extends recent advances in automated PCF calculations to BCC assessment, inheriting key advantages: offering a common and consistent approach, ensuring certifiability and traceability throughout the value chain, and enabling robust scenario analysis and optimization.

The framework will decouple the computational method from value chain data to achieve methodological certification—similar to BASF's SCOTT approach for PCF—rather than relying on product-specific certification. To calculate a product's BCC or any elemental attribute share, the methodology traces atoms through the value chain to their point of origin. The proposed framework derives molecular-level insights by leveraging existing business-focused ERP data, avoiding the need to establish new datasets from scratch.

To achieve these overarching aims, the following objectives are defined:
\begin{itemize}
    \item Identify and curate an industrial value chain case study that encapsulates challenges faced at scale to demonstrate and validate the methodology.
    \item Assess and implement an AI-assisted approach to propose atom mappings of value chain reactions.
    \item Formulate a method for dynamically computing elemental attribute shares of materials based on changing inputs (e.g., feedstock composition, value chain configuration, etc.).
\end{itemize}

Beyond immediate sustainability reporting requirements, this methodology supports broader carbon neutrality goals by facilitating the substitution of fossil-derived inputs with biogenic or recycled alternatives, as discussed by Beer et al. (2025) \cite{beer_forest-based_2025}. Such transparency provides decision-makers with an auditable basis for reducing Scope 1 and Scope 3 emissions while working toward net-zero targets.

A key aspect of this work is the application of existing machine learning models for atom mapping, which significantly reduces the manual burden of tracking chemical reactions across entire product portfolios. By applying state-of-the-art atom mapping algorithms to industrial value chains, we created a modular, operational parameter-agnostic workflow that enables rapid recalculations whenever feedstock compositions or value chain parameters change. This application of existing AI technologies to the BCC calculation problem ensures compliance with TfS requirements while making proactive decarbonization strategies more practical, paving the way for more flexible, transparent, and ultimately greener chemical value chains.

\section{Methodology}\label{sec:methods}
\begin{figure}[h]
    \centering
    \includegraphics[width=0.6\textwidth]{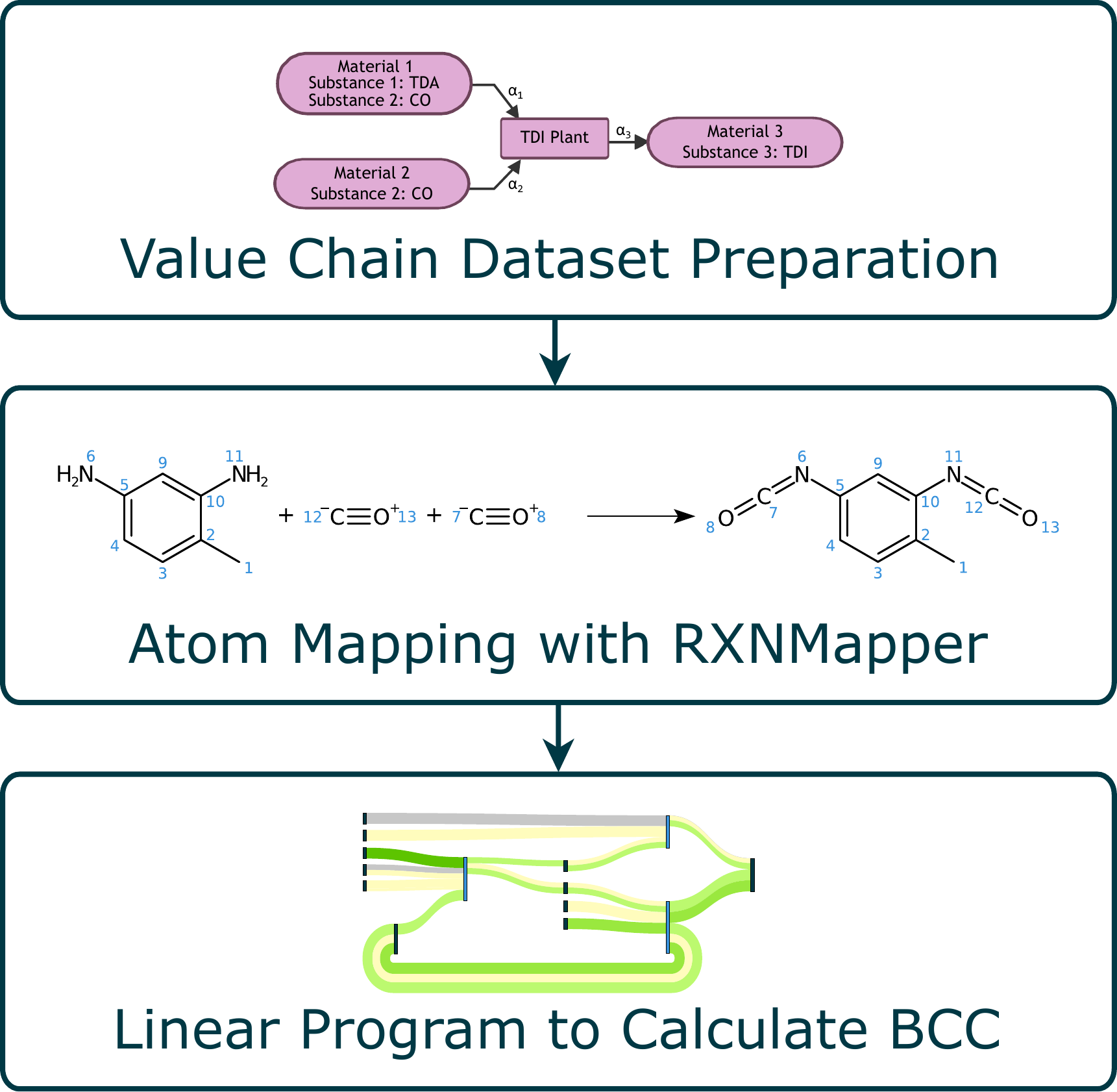}
    \caption{Project methodology and framework for calculating the BCC across a value chain}
    \label{fig:methods}
\end{figure}

We present the CarAT (Carbon Atom Tracing) framework as the solution for determining BCC across industrial value chains. Given a value chain of known inlet attributes, i.e., the fossil and biogenic share of the raw materials, CarAT determines the biogenic carbon share of the products. The three-stage approach is shown in Figure \ref{fig:methods}, and aligns with the project objectives:

\begin{itemize}
    \item \textbf{Industrial Case Studies (Section \ref{sec:2.1})}: Creating a graph representation of the value chain, and pre-processing value chain data.
    \item \textbf{Atom Mapping of Chemical Reactions (Section \ref{sec:2.2})}: Atom mapping chemical reactions using a Chemistry Language model, enabling atom tracing across a production node.
    \item \textbf{Value Chain Model Construction and Optimization (Section \ref{sec:2.3})}: Formulating and solving a linear program to determine the BCC of each substance in the value chain.
\end{itemize}

A realistic example, which is representative of the base-to-speciality chemical industry, is used to demonstrate and verify the framework. For this, an atom mapping machine learning model, RXNMapper, published by IBM \cite{schwaller_extraction_2021}, is leveraged. 

Finally, this framework builds on confidential industrial concepts currently under review in a BASF patent application \cite{citation-key}. Specifically, the concepts and terms bill of materials, bill of substances, and bill of atoms (to be introduced in this section) are included within the scope of the application.

\subsection{Industrial Case Studies}\label{sec:2.1}
An industrial toluene diisocyanate (TDI) value chain of 27 nodes was selected to demonstrate and validate the methodology. Although significantly smaller than a full value chain representing vertically integrated chemical companies such as BASF, Dow, or SABIC, it encapsulates key aspects present in the corresponding large-scale value chains. Furthermore, this work focuses on the transition to carbon neutrality by enabling the replacement of fossil-derived carbon with biogenic or recycled sources. The TDI synthesis pathway represents a realistic candidate for such decarbonization, for example, by substituting the methane used in syngas production with biogas or biomethane \cite{kalinichenko_evaluation_2016}. However, the TDI synthesis lacks a recycle stream—one of the key challenges in value chain calculations. To address this, a smaller butanediol value chain featuring a recycle loop is also examined, demonstrating the framework’s applicability under such complexity.

\subsubsection{Value Chain Graph Topology}
\begin{figure}[h]
    \centering
    \includegraphics[width=0.45\textwidth]{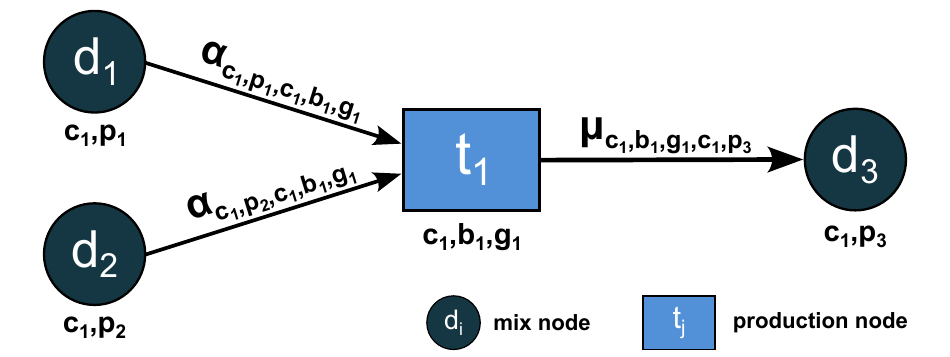}
    \caption{Exemplary schematic of the bipartite directed-graph representation of a value chain. Mix nodes $d_1$, $d_2$, $d_3$, indexed by $(c,p)$ -- here $(c_1,p_1)$, $(c_1,p_2)$, supply feedstock into production node $t_1$, indexed by $(c_1,b_1,g_1)$. Arrows from $d_1$ and $d_2$ to $t_1$ carry input-ratio attributes $\alpha_{c_1,p_1,c_1,b_1,g_1}$ and $\alpha_{c_1,p_2,c_1,b_1,g_1}$, respectively. The arrow from $t_1$ to output mix node $d_3$, indexed by $(c_1,p_3)$, represents the output flow with consumption-mix share $\mu_{c_1,b_1,g_1,c_1,p_3}$.}
    \label{fig:example_vc}
\end{figure}

A value chain can be conceptualized as a bipartite, directed graph, a fundamental mathematical structure in graph theory \cite{asratian_bipartite_1998}. A bipartite graph \( G = (V, E) \) consists of two distinct sets of vertices, \( V = D \cup T \), where \( D \) and \( T \) are disjoint sets (i.e., \( D \cap T = \emptyset \)), and every edge in \( E \) connects a vertex from \( D \) to a vertex in \( T \). 

In the context of the value chain, \( D \) represents virtual tanks, which serve as mix nodes. These virtual tanks are not actual physical containers where chemical reactions occur; rather, they are conceptual nodes introduced to segregate the convergence of a chemical from different sources before entering the actual chemical reactions. Where \( D = \{ d_{1}, d_{2}, \ldots, d_{m} \} \), each \( d_{i} \in D \) signifies a specific virtual tank (e.g.\ \(d_{1},d_{2}\)) in Figure~\ref{fig:example_vc}.

Conversely, \( T =\{ t_1,t_2,\ldots,t_j \} \) denotes production nodes where chemical reactions and transformations occur. Each $t_j \in T$ represents a production step, such as synthesis, separation, formulation, or relabeling (based on ERP data). A triplet $t$ can have one or more input materials that are consumed, and one or more materials that are produced, where \(g\) denotes the main product, and \(p\) represents materials (products, byproducts, and reactants). The given value chain structure is such that one production facility can host more than one triplet \( t_{j} \) -- this can be a consequence of it being a multi-purpose plant, or there being multiple production versions. Each triplet, $t_j$, in the value chain is uniquely identifiable by the ERP data code \((c, b, g)\), where \(c, b, g \) represents the company code, business process, and main product, respectively, e.g., \((c_{1},b_{1},g_{1})\) for \(t_{1}\) in Figure~\ref{fig:example_vc}. Furthermore, each product can be further disaggregated into constituent substances \(s\). Each mix node, $d_i$, is indexed by \((c, p)\)--for instance, \((c_1, p_1)\) is the indentifier for $d_1$ in Figure~\ref{fig:example_vc}. Additionally, \(e\) denotes the chemical element of interest (e.g., carbon) and \(a\) the elemental attribute (e.g., biogenic, fossil, recycled). In this work, \(e\) exclusively refers to carbon, though the CarAT framework is generalizable to other elements.

Edges \(E\) between nodes in \(D\) and \(T\) represent material flows between virtual tanks and production facilities. For edges \((d_{i},t_{j})\in E\), denoted by \(\alpha_{c_i,p_i,c_j,b_j,g_j}\), the attribute \(\alpha\) (input ratio) is defined as the kilograms of material from \(d_{i}\) consumed per kilogram of main output at \(t_{j}\) (see Figure~\ref{fig:example_vc}). Conversely, for edges \((t_{j},d_{i})\in E\), denoted by \(\mu_{c_j,b_j,g_j,c_i,p_i}\), the attribute \(\mu\) (consumption mix share) indicates the fraction of the mixture in \(d_{i}\) originating from \(t_{j}\) (see Figure~\ref{fig:example_vc}).

The value chain is thus modeled as a bipartite directed graph, where edges connect virtual tanks and production nodes. This structure is applied to construct the 27-node TDI value chain. Atom mapping is required only at production nodes, where chemical transformations occur. In contrast, mix nodes (virtual tanks) involve no chemical changes and therefore do not require atom-level tracing.



\subsubsection{TDI Value Chain}
The synthesis pathway in this value chain has been comprehensively detailed in the literature and is protected by patent \cite{tdi_patent}. The primary synthesis route begins with the nitration of toluene, forming 2,4-dinitrotoluene as the major product:
\begin{align}
\text{C}_6\text{H}_5\text{CH}_3 + 2\ \text{HNO}_3 &\rightarrow \text{C}_6\text{H}_3(\text{NO}_2)_2\text{CH}_3 + 2\ \text{H}_2\text{O}
\end{align}
Subsequently, catalytic hydrogenation employing a nickel catalyst reduces the nitro groups of 2,4-dinitrotoluene into amine groups, yielding 2,4-diaminotoluene (TDA):
\begin{align}
\text{C}_6\text{H}_3(\text{NO}_2)_2\text{CH}_3 + 6\ \text{H}_2 &\rightarrow \text{C}_6\text{H}_3(\text{NH}_2)_2\text{CH}_3 + 2\ \text{H}_2\text{O}
\end{align}
In the final stage, TDA undergoes phosgenation to produce TDI alongside hydrochloric acid:
\begin{align}
\text{C}_6\text{H}_3(\text{NH}_2)_2\text{CH}_3 + 2\ \text{COCl}_2 &\rightarrow \text{C}_6\text{H}_3(\text{NCO})_2\text{CH}_3 + 4\ \text{HCl}
\end{align}
This synthesis requires phosgene, derived through a sub-branch of the value chain beginning with the steam reformation of methane-rich natural gas to form syngas, primarily composed of hydrogen and carbon monoxide. Purified carbon monoxide subsequently reacts with chlorine gas, forming the phosgene necessary for the final synthesis step.

\subsubsection{Data Structures for Value Chain Representation}
When working with value chains, it is necessary to create structures other than the graph topology, such as tables listing the node features, such as products, substances, or even atoms present in a given node. This work frequently refers to set datasets that present the value chain information at different levels, defined as follows:
\begin{itemize}    
    \item Bill of materials: a dataset of the recipes at each production node $t_j$, indicating the input ratios of each reactant, defined as the kilograms of material from the duplet consumed per kilogram of main output at the connected triplet, along with the corresponding output ratios of products $p$.
    \item Bill of substances: this adds a further layer of granularity to the bill of materials; it is a dataset of all substances \(s\) for a given production node/set of production nodes.
    \item Bill of atoms on a substance-Level, $\phi_{s'se}$: this is a dataset that designates the share of atoms with attribute \(a\) of a chemical element \(e\) in product substance \(s\) that originates from a reactant substance \(s'\).
    \item Bill of atoms on a material-Level, $\psi_{p's'pse}$: this is a dataset that designates the share of atoms \(a\) of chemical element \(e\) in a product substance \(s\) in product material \(p\) that originates from a reactant substance \(s'\) in reactant material \(p'\); this distinction is particularly important when materials are not pure but are mixtures containing multiple substances.
    \item Consumption mix table: it is a dataset of all \(s\) for a given mix node/set of production nodes.
\end{itemize}

\begin{table}[h]
    \centering
    \renewcommand{\arraystretch}{1.2} 
    \caption{Notation for indices used in the value chain model}
    \begin{tabularx}{0.48\textwidth}{cl}
        \hline
        Index & Description \\
        \hline
        \( a \) & Elemental attribute (e.g., biogenic, fossil, etc.) \\
        \( b \) & Business process, anonymized coding: PLNTb \\
        \( c \) & Company code, anonymized coding: COMPc \\
        \( e \) & Chemical element (e.g., carbon) \\
        \( g \) & Main product, same structure as \( p \) \\
        \( p \) & Product, anonymized coding: PRODp \\
        \( s \) & Substance, represented by SMILES \\
        \hline
    \end{tabularx}
    \label{tab:notation2}
\end{table}

\begin{table}[h]
    \centering
    \renewcommand{\arraystretch}{1.2}    
    \caption{Notation for sets used in graph representation of the value chain}
    \begin{tabularx}{0.48\textwidth}{cl}
        \hline
        Notation & Description \\
        \hline
        \( D \) & Set of mix nodes \\
        \( d \) & A mix/virtual tank node, or duplet \\
        \( T \) & Set of production nodes \\
        \( t \) & A production node, or triplet \\
        \( E \) & Set of value chain edges \\
        \( V \) & Set of all nodes \\
        \hline
    \end{tabularx}
    \label{tab:notation1}
\end{table}
Note that substances are represented by Simplified Molecular Input Line Entry System (SMILES), which is a string notation that allows a user to represent a chemical structure computer-readable format \cite{weininger_smiles_1988}. 

To aid explanation, Table~\ref{tab:bom} and Table~\ref{tab:bos} show an example bill of materials and bill of substances, respectively, for the TDI production node illustrated in Figure~\ref{fig:example}, where Toluene Diamine (TDA) reacts with CO to form TDI. The corresponding bill of atoms is presented in Section~\ref{sec:bcc_calc}. Note that while toluene appears in the node diagram (Figure~\ref{fig:example}), it is omitted from the tables as it was only present in trace amounts below the threshold used in the workflow; substances below this threshold are excluded from further processing to reduce noise and computational burden.
\begin{figure}[H]
    \centering
    \includegraphics[width=0.4\textwidth]{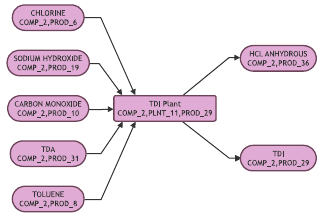}
    \caption{Production node t:COMP2|PLNT11|PROD29, representing the transformation of TDA to TDI}
    \label{fig:example}
\end{figure}
Although not the case here, it is possible to have more than one reactant or product entry with the same substance, i.e., SMILES. This can arise when two different materials share a substance, e.g., if a substance has two different sources. For calculations, the bill of substances can be aggregated such that one entry represents the cumulative amount of each substance for reactants and another entry for products. For example, during atom mapping, the stoichiometric coefficients are estimated for each reaction, necessitating the calculation of moles for each chemical species.

\renewcommand{\arraystretch}{1.1} 
\begin{table}[h!]
    \centering
    \caption{Bill of materials for the TDI production node: t:COMP2|PLNT11|PROD29}
    \begin{tabularx}{0.55\textwidth}{cccc}
        \hline
        Reaction Role & Material & Material Text & Ratio \\ 
        \hline
        Reactant & PROD31 & TDA & 0.53 \\ 
        Reactant & PROD19 & Sodium Hydroxide & 0.04 \\ 
        Reactant & PROD6 & Chlorine & 0.46 \\ 
        Reactant & PROD10 & Carbon Monoxide & 0.52 \\ 
        Product & PROD36 & HCL & 0.63 \\ 
        Product & PROD29 & TDI & 1.16 \\ 
        \hline
    \end{tabularx}

    \label{tab:bom}
\end{table}

\begin{table*}[h]
\centering
\caption{Bill of substances for the TDI production node: t:COMP2|PLNT11|PROD29}
\begin{tabularx}{\textwidth}{@{\extracolsep{\fill}}ccccccc}
\hline
Reaction Role & Material & Material Text & SMILES & Ratio \\
\hline
Product & PROD29 & TDI & Cc1ccc(N=C=O)cc1N=C=O & 1.16 \\
Product & PROD36 & HCL & Cl & 0.56 \\
Product & PROD36 & HCL & O=C=O & 0.02 \\
Product & PROD36 & HCL & [C-]\#[O+] & 0.02 \\
Product & PROD36 & HCL & N\#N & 0.03 \\
Reactant & PROD10 & Carbon Monoxide & [C-]\#[O+] & 0.52 \\
Reactant & PROD6 & Chlorine & ClCl & 0.46 \\
Reactant & PROD19 & Sodium Hydroxide & [Na+].[OH-] & 0.02 \\
Reactant & PROD19 & Sodium Hydroxide & O & 0.02 \\
Reactant & PROD31 & TDA & Cc1ccc(N)cc1N & 0.53 \\
\hline
\end{tabularx}
\label{tab:bos}
\end{table*}

\subsection{Atom Mapping Chemical Reactions}\label{sec:2.2}
Atom mapping, or atom-to-atom mapping (AAM), is a computational chemistry technique that tracks the movement of atoms from the reactants to the products in a chemical reaction \cite{chen_automatic_2013}. In AAM, each atom on the reactant side of the reaction is assigned a unique identifier, which is then transferred to the corresponding atoms on the product side \cite{lin_atom--atom_2022} -- as visualized in Figure~\ref{fig:methods}. This process provides a detailed pathway showing how each atom in the reactants transforms into atoms in the products, crucial for understanding reaction mechanisms and optimizing chemical processes \cite{chen_precise_2024}. Atom mapping finds wide applications across various fields, serving as a fundamental tool for advancing areas such as drug design and other chemical studies \cite{litsa_machine_2019}. It supports the automated identification of reaction centers and the extraction of reaction templates from databases, essential for predicting reaction outcomes \cite{coley_prediction_2017} and training machine learning models used in single-step retrosynthesis \cite{segler_neural-symbolic_2017}.

Determining the BCC of a molecule requires tracing each carbon atom back to its various source materials, distinguishing between fossil-based and renewable sources. This tracing involves retracing the pathway of each carbon atom from reactants through various chemical transformations to the final product in the value chain. Therefore, comprehensive atom-to-atom mapping for each chemical transformation within the value chain is essential. This methodology enables the precise tracking of carbon atoms from inlet materials to final products, ensuring an accurate assessment of BCC.

Several commercially available AAM tools could be used to automate the atom mapping of value chain reactions. However, based on comprehensive benchmarking against popular AAM tools including ChemAxon Automapper, Indigo, RDTool, NameRXN, and RXNMapper, the RXNMapper tool distinguishes itself with an efficient unsupervised-learning transformer model approach \cite{schwaller_extraction_2021}. It achieved the highest accuracy of the AAM tools and was also the second fastest algorithm -- an important factor if such a model were to be deployed at the scale of an entire industrial value chain \cite{lin_atom--atom_2022}.

\subsubsection{RXNMapper from IBM}
RXNMapper is a recent, open-source machine learning (ML) model designed for automatic atom mapping of chemical reactions \cite{schwaller_extraction_2021}. Trained using a self-supervised natural language processing (NLP) approach known as masked language modeling \cite{wang_pre-trained_2023}, RXNMapper learns to predict obscured atoms in reaction SMILES strings, effectively capturing the grammar and complex patterns of chemical reactions.

RXNMapper was selected due to its demonstrated capability to handle intricate reaction details, including stereochemistry and unbalanced reactions, essential for accurately mapping diverse chemical transformations relevant to this study. Benchmark studies report that RXNMapper achieves high accuracy, correctly mapping 99.4\% of a test set comprising 49,000 unbalanced patent reactions sourced from USPTO. Furthermore, it exhibits superior performance compared to other atom mapping tools such as Indigo \cite{noauthor_indigo_nodate} and Mappet \cite{wojciech1_automatic_2019}, providing the fastest inference times at 7.7 ms per reaction using a GPU \cite{schwaller_extraction_2021}.

A detailed overview of RXNMapper, including its architecture based on transformer neural networks, training methodology, and performance evaluation, is provided in the \nameref{sec:SI}.

\subsubsection{Atom Mapping Workflow: TDI Production Node}
\begin{figure*}[t]
    \centering
    \includegraphics[width=\textwidth]{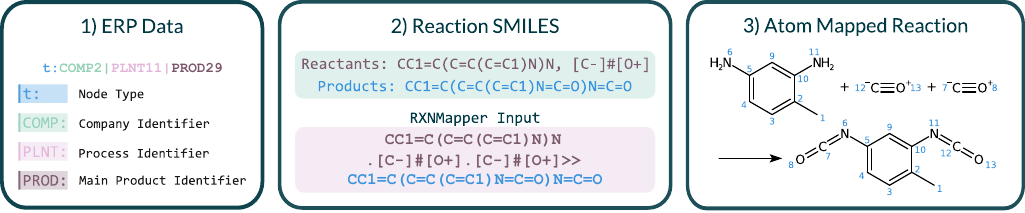}
    \caption{Atom mapping workflow: convert ERP data to molecular structures, construct reaction SMILES, apply RXNMapper to generate atom-mapped SMILES, then convert to bill of atoms format}
    \label{fig:example_aam}
\end{figure*}

Figure~\ref{fig:example_aam} illustrates the atom mapping workflow, which transforms ERP data into atom-mapped molecular structures through a three-step process. The TDI production node serves as an example to demonstrate this workflow.

\noindent \textbf{Step 1: ERP Data Preprocessing}

The workflow begins by preprocessing ERP data to identify chemical species and convert them into molecular structures. This process generates a bill of substances for the node (Table~\ref{tab:bos}), which lists all relevant species as canonical SMILES strings. These molecular structures form the foundation for subsequent analysis.

\noindent \textbf{Step 2: Reaction SMILES Construction}

In the second stage, reaction SMILES are constructed—a linear string notation that encodes chemical transformations from input substances to output products. The standard reaction SMILES syntax follows the format: 
\texttt{[reactants] > [reagents] > [products]}. 

This work adopts a simplified approach: all input substances are included in the reactant section, while the reagent section remains empty. This simplification streamlines the parsing process without affecting the results, as RXNMapper only annotates atoms in reactants and products. This generic structure enables efficient computational analysis, facilitating tasks such as reaction prediction, optimization, and data mining~\cite{zhong_root-aligned_2022}.

Since RXNMapper requires reaction SMILES with only one product substance, multiple reaction SMILES strings must be constructed for nodes with multiple products. For a production node with $j$ reactant substances and $k$ products, $k$ reaction SMILES strings are generated—each containing the same set of reactants but differing in the product species, as shown in Equation~\ref{eq:stoich}.
The stoichiometric coefficients are estimated using the mole quantity of each substance ($n_s$), calculated by Equation~\ref{eq:mols}:
\begin{equation}
\frac{n_1}{n_p} \text{reactant}_1 + \frac{n_2}{n_p} \text{reactant}_2 + \cdots + \frac{n_j}{n_p} \text{reactant}_j \rightarrow \text{product}
\label{eq:stoich}
\end{equation}
\begin{equation}
n_s = \frac{1}{M_s} \sum_p \alpha_{p} \cdot \lambda_{ps}
\label{eq:mols}
\end{equation}
where:
\begin{itemize}
    \item $\lambda_{ps}$ is the mass ratio of substance $s$ in product $p$.
    \item $\alpha_p$ is the input ratio (kg of material from duplet consumed per kg of main output at connected triplet).
    \item $M_s$ is the molar mass of substance $s$.
\end{itemize}

\noindent \textbf{Step 3: Atom Mapping and Bill of Atoms Generation}\\
The constructed reaction SMILES are passed to the RXNMapper model, which returns atom-mapped reaction SMILES. Figure~\ref{fig:methods} displays both the unmapped reaction SMILES and a visualization of the mapped reaction output for the TDI production node.
For enhanced interpretability, the mapped reactions can be visualized using CDK Depict \cite{noauthor_cdk_nodate} or RDKit \cite{rdkit}, as shown in the third stage of Figure~\ref{fig:example_aam}.
To calculate the BCC, the atom mapping must be translated into a ``bill of atoms'' format \cite{citation-key}. The atom mapping directly yields the substance-level atom bill, denoted $\phi_{s'se}$, which represents the share of atoms of chemical element $e$ in output substance $s$ that originated from input substance $s'$. Table~\ref{tab:boa_node} presents the complete bill of atoms for the TDI production node. Note that for this framework, the bill of atoms is only required for carbon-containing materials.
\renewcommand{\arraystretch}{1.1} 
\begin{table*}[h!]
    \centering
    \caption{Bill of atoms for TDI production node: t:COMP2|PLNT11|PROD29}
    \small
    \begin{tabularx}{\textwidth}{@{\extracolsep{\fill}}lllllll}
    \hline
    Reactant Material & Reactant SMILES & Product Material & Product SMILES & Element & Atom Count & Atom Share \\
    \hline
    PROD10 & [C-]\#[O+] & PROD36 & O=C=O & O &  1 & 0.50 \\
    PROD10 & [C-]\#[O+] & PROD36 & O=C=O & C & 1 & 1.00 \\
    PROD31 & Cc1ccc(N)cc1N & PROD29 & Cc1ccc(N=C=O)cc1N=C=O & C & 7 & 0.78 \\
    PROD31 & Cc1ccc(N)cc1N & PROD29 & Cc1ccc(N=C=O)cc1N=C=O & H & 6 & 1.00 \\
    PROD31 & Cc1ccc(N)cc1N & PROD29 & Cc1ccc(N=C=O)cc1N=C=O & N &  2 & 1.00 \\
    PROD10 & [C-]\#[O+] & PROD29 & Cc1ccc(N=C=O)cc1N=C=O & O & 2 & 1.00 \\
    PROD10 & [C-]\#[O+] & PROD29 & Cc1ccc(N=C=O)cc1N=C=O & C & 2 & 0.22 \\
    PROD6 & ClCl & PROD36 & Cl & Cl & 1 & 1.00 \\
    PROD6 & ClCl & PROD36 & Cl & H & 1 & 1.00 \\
    PROD31 & Cc1ccc(N)cc1N & PROD36 & N\#N & N & 2 & 1.00 \\
    PROD19 & O & PROD36 & O=C=O & O & 1 & 0.50 \\
    PROD10 & [C-]\#[O+] & PROD36 & [C-]\#[O+] & O & 1 & 1.00 \\
    PROD10 & [C-]\#[O+] & PROD36 & [C-]\#[O+] & O & 1 & 1.00 \\
    \hline
\end{tabularx}
\label{tab:boa_node}
\end{table*}

\subsection{Model Construction and Optimization}\label{sec:2.3}
With the atom mapping procedure complete, the necessary data is available to calculate the BCC for a given production node. However, to apply this approach to a value chain involving multiple interconnected mix and production nodes as per the third project objective, a system of equations must be formulated. Solving this system will provide the BCC for every material and substance throughout the value chain.

This section will first present a detailed example of calculating the BCC for a single production node. Subsequently, the value chain system will be defined, and a suitable method for solving the system will be selected and discussed. In addition to the graph notation introduced in Table \ref{tab:notation1}, Table \ref{tab:notation2} defines indices required for methodology.


\subsubsection{BCC Calculation for TDI Production Node}\label{sec:bcc_calc}

From the atom mapping section of the workflow, the bill of atoms on a substance level, $\phi_{s'se}$, is calculated, which denotes the share of atoms of chemical element $e$ in outlet substance $s$ that originate from the inlet substance $s'$. However, the atom bill is required on a material level, $\psi_{p's'pse}$ for instances where the materials are not pure, i.e., contain more than one substance. The material-level atom bill denotes the share of chemical element $e$ in outlet substance $s$, in outlet material $p$ that originates from the inlet substance $s'$ within inlet material $p'$. Equation \ref{eq:mat_boa} shows how the material-level atom bill can be calculated given the substance-level atom bill, where $\lambda_{p's'}$ represents the mass fraction of inlet substance $s'$ in inlet material $p'$.
\begin{equation}
\begin{aligned}
    \psi_{p's'pse} = \frac{\alpha_{p'} \cdot \lambda_{p's'}}{\sum_{p'} \alpha_{p'} \cdot \lambda_{p's'}} \phi_{s'se}
\end{aligned}
\label{eq:mat_boa}
\end{equation}
With the material-level atom bill in place, two further equations are required to fully define a system to determine the BCC of a value chain system. These equations focus on calculating the elemental attribute share~$\beta$. Equation \ref{eq:beta_trip} is specific to calculating the share of attribute $a$ (e.g., fossil, biogenic, etc) of chemical element~$e$ in substance~$s$ in material~$p$ within the production node~($c,b,g$). It does so by summing the attribute contributions from each incoming mix node denoted~($c',p'$), weighted by the material-level atom bill~$\psi_{p's'pse}$.
\begin{equation}
\begin{aligned}
    \beta_{cbgpsea} = \sum_{p's'}\psi_{p's'pse} \: \beta_{cp's'ea}
\end{aligned}
\label{eq:beta_trip}
\end{equation}
Equation \ref{eq:beta_dup} calculates the attribute share $a$ of element $e$ in substance $s$ in material $p$ within the mix node $(c,p)$. It does so by summing the attribute contributions from each incoming production node, denoted ($c',b',g'$), weighted by the consumption mix share (i.e., how much from the mix node comes from each production node), $\mu_{c'b'g'cp}$.
\begin{equation}
\begin{aligned}
    \beta_{cpsea} = \sum_{c'b'g'}\mu_{c'b'g'cp} \: \beta_{c'b'g'psea}
\end{aligned}
\label{eq:beta_dup}
\end{equation}


\subsubsection{Example: One Node Calculation}\label{sec:one_node_calc}
\begin{figure}[H]
    \centering
    \includegraphics[width=0.45\textwidth]{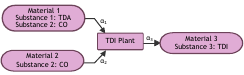}
    \caption{Example of a TDI production node with impure inlet materials, corresponding to triplet \( t = (c, b, g) \)}
    \label{fig:node_example}
\end{figure}

This example demonstrates the BCC calculation for a production node \( t = (c, b, g) \), where the main one-product reaction for the formation of TDI is:
\begin{center}
    \ce{C7H10N2 + 2CO -> C9H6N2O2}
\end{center}

TDA, carbon monoxide (CO), and TDI are represented as substances 1, 2, and 3, respectively. In this case, CO is present in both input materials:

\begin{itemize} 
    \item Material 1: 80\% TDA, 20\% CO → $\lambda_{11} = 0.8$, $\lambda_{12} = 0.2$ 
    \item Material 2: 100\% CO → $\lambda_{22} = 1$ 
\end{itemize}

Let the corresponding input ratios be:
\[
\alpha_1 = \frac{20}{60}, \quad \alpha_2 = \frac{40}{60}
\]

The substance-level atom bill, calculated using the atom mapper tool, $\phi_{s'se}$, is: 
\[\phi_{13C} = \frac{7}{9}, \quad \phi_{12C} = \frac{2}{9}\]

Since substance 1 (TDA) is only present in material 1, the material-level and substance-level atom bills are equivalent:
\[\psi_{1133C} = \phi_{13C} = \frac{7}{9}\]

However, for CO, which is split across both materials, the material-level atom bills are calculated using Equation \ref{eq:mat_boa}:
\begin{align*}
    \psi_{1233C} &= \frac{\alpha_1 \lambda_{12}}{\alpha_1 \lambda_{12} + \alpha_2 \lambda_{22}} \cdot \phi_{12C} \quad = \frac{\frac{20}{60} \cdot 0.2}{\frac{20}{60} \cdot 0.2 + \frac{40}{60} \cdot 1} \cdot \frac{2}{9} = \frac{2}{99} \\
    \psi_{2233C} &= \frac{\alpha_2 \lambda_{22}}{\alpha_1 \lambda_{12} + \alpha_2 \lambda_{22}} \cdot \phi_{12C} \quad = \frac{\frac{40}{60} \cdot 1}{\frac{20}{60} \cdot 0.2 + \frac{40}{60} \cdot 1} \cdot \frac{2}{9} = \frac{20}{99}
\end{align*}

Assuming material 1 is entirely fossil-derived and material 2 is entirely biogenic, let \( A \) be the set of elemental attributes considered (e.g., fossil, biogenic), and in this example, let \( x \in A \) denote biogenic carbon:

\[\beta_{c11Cx}, \: \beta_{c12Cx} = 0, \quad \beta_{c22Cx} = 1\]

The BCC for this TDI node is then computed as:
\[\beta_{c33Cx} = \beta_{c11Cx} \cdot \psi_{1133C} + \beta_{c12Cx} \cdot \psi_{1233C} + \beta_{c22Cx} \cdot \psi_{2233C}\]

\[\beta_{c33Cx} = 0 \cdot \frac{7}{9} + 0 \cdot \frac{2}{99} + 1 \cdot \frac{20}{99} = \frac{20}{99}\]

\subsubsection{Selection of a Linear Program Approach}
Section~\ref{sec:one_node_calc} demonstrated the procedure for calculating the BCC for a single product in a production node. However, in a value chain, the BCC is dependent on the BCC of all preceding nodes. Additionally, due to the presence of recycle streams, it would also be infeasible to sequentially determine the BCC by carrying out the calculation on a one-node-at-a-time basis. Hence, to compute the BCC across the value chain, a linear system of equations is defined, with Equations \ref{eq:beta_trip} \& \ref{eq:beta_dup} at the core of this system. A Linear Program (LP) is chosen to solve the system of equations -- a practical choice that enables the problem to be posed as a slack minimization. A linear optimization model is characterized by the following criteria: only continuous variables, a single linear objective function, and only linear equality or inequality constraints. 

The BCC system can be formulated as a feasibility problem, wherein the objective is not to optimize a particular function, but rather to identify values for the elemental attribute shares of each substance that satisfy a set of constraints. In practice, however, industrial datasets often contain inconsistencies or incomplete information that may render the constraint set infeasible. To accommodate such cases, slack variables are introduced, allowing for controlled violations of the constraints. This enables the model to yield a solution even when exact feasibility is not attainable, while also quantifying the extent of any deviations.

In this formulation, the objective function is defined to minimise the total system slack. This drives the solution towards that of the original feasibility problem under the assumption of fully consistent and accurate data. Moreover, the magnitude and location of slack values provide diagnostic insight by identifying specific constraints where data limitations are most pronounced. As such, the use of slack variables offers both computational robustness and practical interpretability, making the approach particularly valuable in industrial contexts where data uncertainty is common.

The LP formulation was implemented using the Python MIP package \cite{noauthor_python-mip_nodate}, using the CBC (COIN-OR branch and cut) solvers \cite{forrest_coin-orcbc_2023} -- as it is open-source and suitable for LPs.

\subsubsection{Linear Program Formulation}
    \begin{equation}\label{eq:lp}
    \begin{alignedat}{4}
      &\min_{\mathbf{\theta}}\;
        &&\sum_{c,b,g,p,s,e}(z_{cbgpse}-q_{cbgpse})
        +\sum_{c,p,s,e}(z_{cpse}- && q_{cpse})
     && \text{(a)} \\
    &\text{s.t.} \\[-0.4cm]
    & && \beta_{cbgpsea} = \sum_{p's'}\psi_{p's'pse}\,\beta_{cp's'ea}, 
    && \forall c,b,g,p,s,e,a \quad && \text{(b)} \\[4pt]
    & && \beta_{cpsea} = \sum_{c'b'g'}\mu_{c'b'g'cp}\,\beta_{c'b'g'psea}, 
    && \forall c,p,s,e,a &&\text{(c)} \\[4pt]
    & && \sum_a \beta_{cbgpsea}-z_{cbgpse}-q_{cbgpse}=1, 
    && \forall c,b,g,p,s,e &&\text{(d)} \\[4pt]
    & && \sum_a \beta_{cpsea}-z_{cpse}-q_{cpse}=1, 
    && \forall c,p,s,e &&\text{(e)} \\[4pt]
    & && \beta_{cbgpsea} \in [0,1], 
    && \forall c,b,g,p,s,e,a &&\text{(f)} \\[4pt]
    & && \beta_{cpsea} \in [0,1], 
    && \forall c,p,s,e,a &&\text{(g)} \\[4pt]
    & && z_{cbgpse} \in \mathbb{R}^+, \quad q_{cbgpse} \in \mathbb{R}^-, 
    && \forall c,b,g,p,s,e &&\text{(h)} \\[4pt]
    & && z_{cpse} \in \mathbb{R}^+, \quad q_{cpse} \in \mathbb{R}^-, 
    && \forall c,p,s,e &&\text{(i)}
    \end{alignedat}
    \end{equation}

\vspace{0.3cm}

N.B. \(c'\) denotes the inlet company code, whereas \(c\) denotes the outlet company code, and $\boldsymbol{\theta}$ is the set of decision variables $\beta_{cbgpsea}$, $\beta_{cpsea}$, $z_{cbgpse}$, $q_{cbgpse}$, $z_{cpse}$, and $q_{cpse}$ for all company codes~$c$, business processes~$b$, main products~$g$, materials~$p$, substances~$s$, chemical elements~$e$, and elemental attributes~$a$.

The objective function (\ref{eq:lp}a) minimizes slack across the entire value chain, encouraging efficient use or elimination of slack variables. Table~\ref{tab:variables_parameters} summarizes the decision variables and parameters used in the LP. Slack variables are denoted $z$ (positive) and $q$ (negative), with subscripts indicating context: $z_{cpse}$ and $q_{cpse}$ for duplets, $z_{cbgpse}$ and $q_{cbgpse}$ for triplets. 

\begin{table}[h]
    \centering
     \caption{Decision variables, slack variables, and parameters}
     \renewcommand{\arraystretch}{1.2}
     \begin{tabularx}{0.7\textwidth}{>{\centering\arraybackslash}m{2.0cm} m{8.8cm}}
         \hline
         Notation & Description \\
        \hline
        \( \beta_{cbgpsea} \) & Fraction of elemental attribute \( a \) of chemical element \( e \) in substance \( s \), material \( p \), at production node \( (c, b, g) \). \\
        \( \beta_{cpsea} \) & Fraction of elemental attribute \( a \) of chemical element \( e \) in substance \( s \), material \( p \), at mix node \( (c, p) \). \\
         \( z_{cbgpse} \) & Positive slack variable for chemical element \( e \) in substance \( s \), material \( p \), at production node \( (c, b, g) \). \\
         \( q_{cbgpse} \) & Negative slack variable for chemical element \( e \) in substance \( s \), material \( p \), at production node \( (c, b, g) \). \\
         \( z_{cpse} \) & Positive slack variable for chemical element \( e \) in substance \( s \), material \( p \), at mix node \( (c, p) \). \\
         \( q_{cpse} \) & Negative slack variable for chemical element \( e \) in substance \( s \), material \( p \), at mix node \( (c, p) \). \\
        \( \mu_{c'b'g'cp} \) & Mix node share, i.e., the fraction of a virtual tank \( (c,p) \) sourced from a production node \( (c',b',g') \). \\
         \( \psi_{p's'pse} \) & Bill of atoms, i.e., the fraction of chemical element \( e \) in substance \( s \) in product \( p \), sourced from substance \( s' \) in product \( p' \). \\
         \hline
     \end{tabularx}
     \label{tab:variables_parameters}
 \end{table}

\subsection*{Inlet Conditions}
Solving this formulation is dependent on the inlet nodes entering the value chain subgraph having known values of $\beta_{cpsea}$. To express this, we define a set of inlet mix nodes, denoted by $D_0$. These are nodes in the set of all mix nodes $D$ that have no incoming edges. We formalize this as:
\begin{equation}
    D_0 = \{d_i \in D \mid \delta^-(d_i) = \emptyset\}
\end{equation}
Here, $d_i$ refers to a particular mix node in $D$, and $\delta^-(d_i)$ denotes the set of inlet edges into node $d_i$; if this set is empty, then $d_i$ is an inlet node. For simplicity in formulation, a structural assumption is imposed: the value chain must start and end with mix nodes. Hence, the values of $\beta_{cpsea}$ at the inlet mix nodes in $D_0$ must be specified and set to a known constant $h$:
\begin{equation}
    \beta_{c_{d_i} p_{d_i} sea} = h \quad \forall \; d_i \in D_0, \; \forall \; s, e, a
\end{equation}

\subsection*{Constraints}
\begin{itemize}
    \item Constraints (\ref{eq:lp}b) and (\ref{eq:lp}c) define elemental attribute share equations for production and mix nodes, respectively. 
    \item Equations (\ref{eq:lp}d) and (\ref{eq:lp}e) ensure that for any chemical substance, the sum of carbon attributes (including fossil and biogenic) equals one, incorporating slack variables.
    \item Equations (\ref{eq:lp}f) and (\ref{eq:lp}g) set bounds on elemental attribute shares for production and mix nodes, ensuring \( \beta \) values are non-negative and do not exceed one.
    \item Equations~(\ref{eq:lp}h) define non-negative bounds for positive and negative slack variables in production nodes. Equation~(\ref{eq:lp}i) establishes analogous bounds for mix nodes.
\end{itemize}

\begin{figure*}[!htbp]
    \centering
    \includegraphics[width=0.9\textwidth]{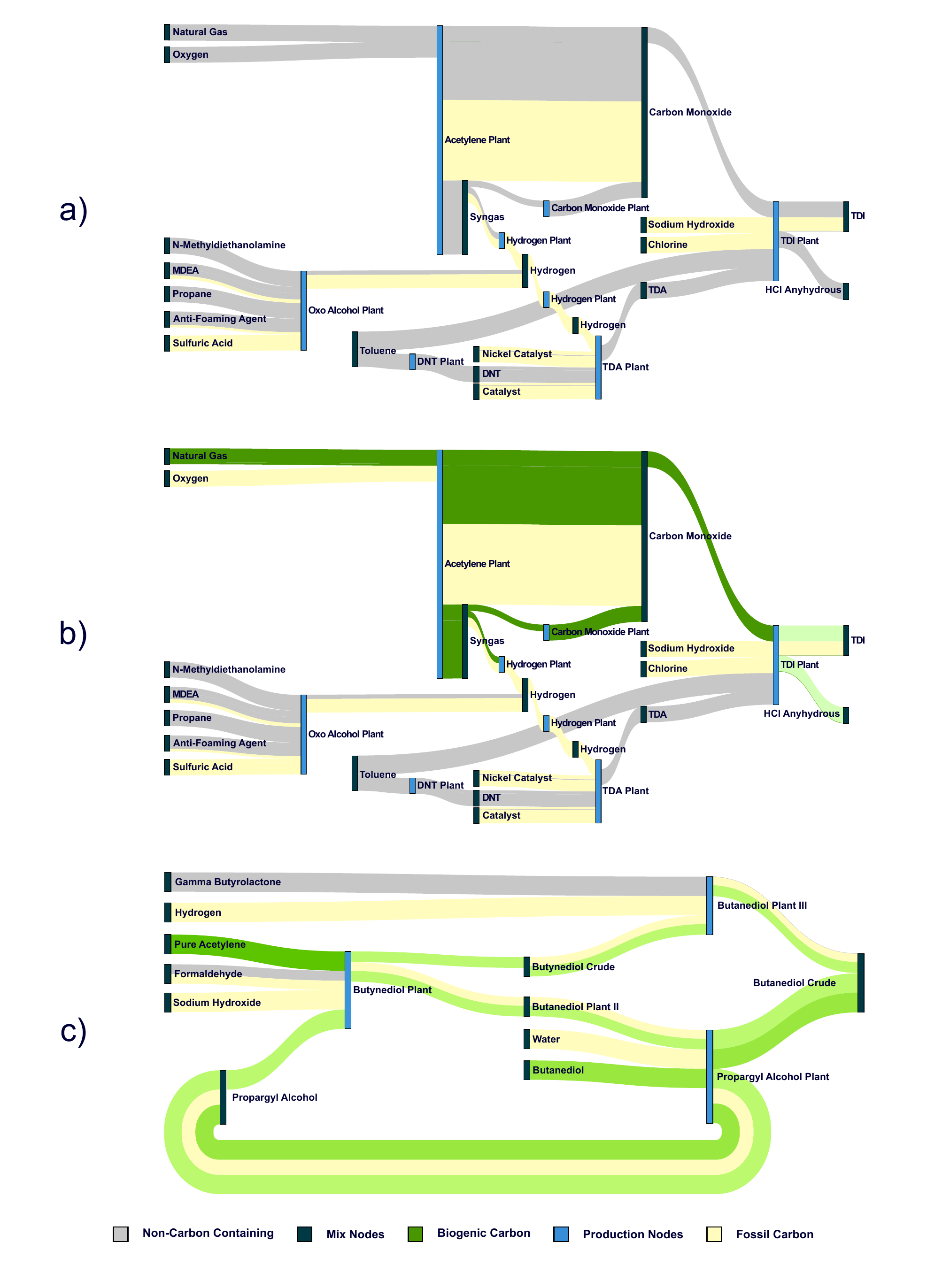}
    \caption{Sankey diagrams for (a) Base Case, (b) Case 1, and (c) Case 2. Node colors: dark blue = mix, light blue = production. Link colors: gray = fossil carbon, green gradient = biogenic carbon (intensity of green is proportional share), yellow = non-carbon streams. Overall BCCs: 0 \%, 22 \%, 38 \%, respectively.}
    \label{fig:three_sankeys}
\end{figure*}

\section{Results}
Building on the methodology presented in Section \ref{sec:methods}, we apply CarAT to an industrial TDI value chain and, to demonstrate recycle-stream handling, a butanediol value chain. These examples were selected for their industrial relevance and contrasting topologies. Three scenarios are analyzed:

\begin{itemize}
    \item \textbf{Base Case -- \nameref{sec:SI}}: TDI value chain with entirely fossil-derived carbon inputs. TDI product BCC = 0\%. 
    \item \textbf{Case 1 -- Section \ref{sec:3.2}}: TDI value chain with 100\% biogenic natural-gas feed. TDI product BCC = 22\%.
    \item \textbf{Case 2 -- Section \ref{sec:3.3}}: BDO value chain featuring a recycle stream; 75\% biogenic acetylene and 50\% biogenic BDO inlet. Butanediol product BCC = 38\%.
\end{itemize}

The three-stage CarAT methodology is fully implemented in a Python package and demonstrated here through three worked scenarios. For each case, the linear program results are visualized as a Sankey diagram, which clearly conveys the carbon attribute flows and the bipartite graph structure of the value chain. Importantly, the objective function—i.e., the total system slack—for all three scenarios is numerical zero, indicating successful LP convergence. The complete analysis, including value chain data, solver setup, and visualization scripts, is available in a public GitHub repository \cite{pajak_optimal-pse-labcarat_2025}, ensuring full reproducibility and enabling others to apply CarAT to new chemical systems. The base case scenario, being structurally simple and yielding a zero BCC by design, is discussed in the Supporting Information (Section~\ref{sec:SIbasecase}).

To better visualize the flow of elemental attributes, particularly the flow of biogenic carbon, a color-coding scheme was implemented. In the diagram, dark blue bars represent mix nodes and light blue bars represent production nodes. Pale yellow links represent the flow of non-carbon-containing compounds, which are still included to provide a full picture of the value chain. To differentiate between carbon and non-carbon-containing compounds, a link is shown for each substance (SMILES) transferred between the two nodes. The thickness of the links entering the blue mix nodes is proportional to the consumption mix share of that substance $\mu_{c'b'gcp}$. Similarly, the thickness of the links entering the red production nodes is proportional to the input ratio of that substance at that node $(c,b,g)$.

Importantly, fossil carbon is shown in gray, and green links represent biogenic carbon, with darker green signifying a larger share of biogenic carbon and paler green representing a smaller share. As expected, the carbon in the resulting TDI product stream from this value chain is 100\% fossil-based.

\subsection{Case 1: Biogenic Methane Feedstock}\label{sec:3.2}

Case 1 (Figure~\ref{fig:three_sankeys}b) features a 100\% biogenic natural gas feedstock, shown in dark green. The biogenic methane and ethane enter a production node to form carbon monoxide. The resulting carbon monoxide enters the TDI production node, where it reacts with chlorine to form phosgene, which subsequently reacts with TDA to yield TDI, alongside a byproduct of anhydrous hydrochloric acid.

Since TDI consists of nine carbon atoms—two sourced from the 100\% biogenic carbon monoxide and seven from the fossil-derived TDA—the resulting TDI has a BCC of 22\%, indicated by a lighter shade of green. This scenario demonstrates the capability to calculate the BCC of a product when using mixed sources of carbon feedstock in its synthesis.

\subsection{Case 2: Butanediol Value Chain with a Recycle Stream}\label{sec:3.3}

In Case 2 (Figure~\ref{fig:three_sankeys}c), based on the butanediol value chain, all carbon is fossil-derived except for a 75\% biogenic pure acetylene feedstock and a 50\% biogenic butanediol stream. The 75\% biogenic acetylene reacts with two equivalents of fossil-derived formaldehyde, producing butynediol with a BCC of 38\%. Following hydrogenation, the resulting butanediol retains this BCC of 38\%. In the recycle stream, two carbon-containing substances are present: 50\% biogenic butanediol and propargyl alcohol, a byproduct from the butynediol synthesis.

The recycle stream adds complexity and reflects the reality of industrial value chains where such flows are common. The framework correctly apportions biogenic carbon through the recycle path, demonstrating its robustness to cyclic topologies.

\section{Discussion and Limitations}\label{sec:lim}

The results above confirm that CarAT quantifies the BCC of value chains ranging from a simple linear TDI route to a cyclic BDO system. Two practical constraints now merit discussion: (i) atom-mapping accuracy and (ii) data completeness.

\subsection{Atom-mapping accuracy}

RXNMapper delivers high overall accuracy\cite{schwaller_unsupervised_2020,lin_atom--atom_2022}, yet occasional mis-mappings are inevitable. We envisage a two-tier mitigation strategy:

\begin{enumerate}
  \item \textbf{Domain expert supersession of automated mapping.} In industrial deployment, chemists can flag and correct mis-mapped reactions during data onboarding, ensuring critical pathways are traced correctly. Having some form of confidence in the atom mapping--such as that provided by the more recent LocalMapper \cite{chen_precise_2024}--could help guide targeted spot checks.
  \item \textbf{Reaction-class-specific fine-tuning.} If systematic errors are observed for a given reaction class, additional training on that subset can be leveraged to improve accuracy—see Section~\ref{sec:future_work}.
\end{enumerate}

RXNMapper’s default token limit (512) was sufficient for the value chains analyzed in this work. However, as reactions become more complex downstream in an industrial value chain, the impact of impurities, solvents, and other non-reacting compounds could make the size limitation more significant. To address this, initial efforts could involve using RXNMapper without considering stoichiometry, as the model has demonstrated good accuracy even with unbalanced reactions \cite{schwaller_learning_2021}. Furthermore, careful curation of the reaction SMILES might be beneficial, such as removing non-essential compounds while preserving the core chemical reaction and essential reactants.

\subsection{Data Completeness}

A more fundamental bottleneck is missing or non-standardized molecular identifiers. CarAT requires a canonical SMILES for every carbon-containing species; where none exists, the framework is unable to trace atoms. Conversion from CAS numbers could alleviate part of this gap, but a persistent lack of digital representations for proprietary intermediates remains a barrier to full-chain BCC accounting. Expanding internal chemical inventories or adopting open-identifier policies will therefore be as important as future algorithmic improvements.

\section{Conclusion}
This research addresses an urgent challenge facing the chemical industry: the need for a scalable, generalizable, and dynamic methodology to calculate the BCC of chemical products. As the TfS consortium prepares to mandate BCC reporting by 2026, companies must be able to track the origin of carbon atoms across complex and evolving value chains. Existing methods are static and product-specific, limiting their use in an industry characterized by changing feedstocks, recycling, and distributed manufacturing.

To meet this challenge, we have developed CarAT, a framework that integrates enterprise-level data with a chemistry language model and linear optimization to automate carbon tracing at scale. By mapping carbon atoms from feedstocks through each stage of production and solving for BCC via a linear program, CarAT decouples data from methodology. This structure allows manufacturers to obtain methodological certification, enabling automatic recalculations as operational parameters evolve.

Validation was carried out across three scenarios that increased in complexity: a fossil-only TDI value chain, the same chain with a 100\% biogenic natural gas inlet, and a butanediol value chain incorporating a recycle stream and partial biogenic inputs. These scenarios tested the framework’s capacity to adapt to key industrial features, including mixed feedstocks and recycling loops, with results confirming both robustness and scalability. Having demonstrated its robustness through these case studies, CarAT is now being implemented within BASF to proactively trace the carbon attributes of its product portfolio. This industrial adoption affirms CarAT’s potential for large-scale deployment and highlights the chemical sector’s pressing demand for scalable carbon-tracing solutions.

Further to supporting compliance with sustainability reporting requirements, CarAT enables informed decision-making for decarbonization. By tracing biogenic carbon content and quantifying the impact of raw material composition on product-level BCC, it facilitates the substitution of fossil-based carbon with biogenic alternatives. This, in turn, guides internal decisions and enhances value chain transparency, enabling Scope 3 emissions reductions through more informed upstream raw material choices.

By clearly aligning with key Green Chemistry principles, particularly around substituting fossil feedstocks with biogenic or recycled sources, CarAT also offers a concrete step toward net-zero goals. Its capacity for rapid recalculation facilitates real-time adjustments in sourcing and operational strategies, making sustainable innovation both more transparent and more feasible across the chemical industry.

\subsection{Future Work}\label{sec:future_work}
Future work will proceed along three key avenues. First, improving the accuracy of the atom mapping is essential for broader adoption. Although RXNMapper performed well in validation, its limitations become more apparent with long or complex SMILES strings. As the framework is applied to larger and more diverse industrial datasets, identifying specific reaction classes where mapping accuracy deteriorates (or a wider context window is required) will be an important step. These can then be addressed by fine-tuning RXNMapper on appropriately curated, related reaction sets, expanding its context window, or developing a custom transformer-based model.

Second, the linear program underlying CarAT is generalizable beyond carbon. With suitable data, the same methodology could be extended to trace other elemental attributes such as nitrogen, recycled content, or toxic elements. This would enable more comprehensive sustainability assessments across chemical value chains.

Third, future work will explore the inverse optimization problem: how to allocate biogenic raw materials across a value chain to achieve a desired biogenic carbon content in the final product. This has clear relevance for setting and meeting emissions targets across value chains.

Taken together, these developments will support the evolution of CarAT into a more robust and flexible tool for sustainability analysis within value chains, with the potential to assist industry in its transition towards net zero.

\section{Acknowledgements}
Financial support provided by BASF SE and EPSRC CDT (EP/S023232/1) is acknowledged. The authors thank the BASF SCOTT team and the OptiML group at Imperial College London for their support and discussions. The authors also express their gratitude to Friedrich Hastedt for their valuable feedback and proofreading assistance.

\section{Code and Data Availability}
The CarAT Python package and example value chain datasets used in this work are openly available on Zenodo at \href{https://doi.org/10.5281/zenodo.16849843}{doi:10.5281/zenodo.16849843}.

\printbibliography
\section{\hyperref[sec:SIbasecase]{Supplementary Information}}\label{sec:SI}
\subsection*{TDI Value Chain Case Study}

\begin{figure}[h]
    \centering
    \includegraphics[width=\linewidth]{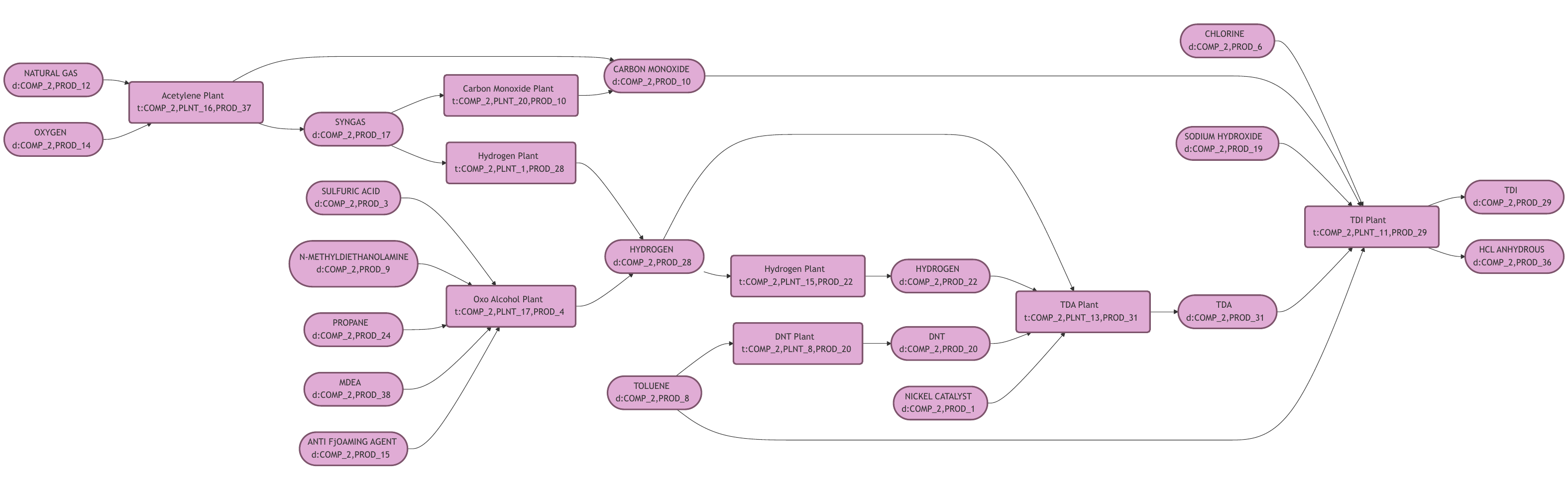}
    \caption{Bipartite directed graph of the toluene diisocyanate value chain case study, comprising 27 nodes}
    \label{fig:full_vc_graph}
\end{figure}

\subsection*{Overview of RXNMapper}\label{sec:SI_rxnmapper}

Natural Language Processing (NLP) is a subfield of machine learning that enables computers to understand, interpret, and generate human language. NLP includes tasks such as text analysis, language translation, and question answering \cite{jurafsky_speech_2009}. Within chemistry, NLP is facilitated by SMILES notation, a text-based representation that encodes chemical structures into strings, thus enabling computational applications such as property prediction and chemical compound generation \cite{ozturk_exploring_2020}.
\\
RXNMapper employs a self-supervised NLP technique called masked language modeling, trained on 2.8 million chemical reactions \cite{wang_pre-trained_2023}. In this approach, certain atoms within reaction SMILES strings are obscured, prompting the model to predict the missing atoms based on contextual clues from surrounding atoms. This method allows RXNMapper to implicitly learn chemical grammar and complex reaction patterns from the data itself.

\subsubsection*{Transformer Neural Networks}
Transformer neural networks, introduced in the landmark paper ``Attention Is All You Need'' \cite{vaswani_attention_2023}, have emerged as a state-of-the-art technique in NLP. Transformers differ significantly from traditional Recurrent Neural Networks (RNNs) through their use of self-attention, enabling simultaneous processing of input sequences and effectively handling long-range dependencies.
\\
RXNMapper utilizes the ALBERT (A Lite BERT) architecture, a variant of the widely used BERT model known for bidirectional context processing. ALBERT shares weights across layers during training, resulting in a smaller model size that retains consistent functionality across different layers and inputs \cite{lan_albert_2020, schwaller_learning_2021}. This capability is particularly valuable for accurately modeling complex chemical reactions.
\\
Performance assessments indicate RXNMapper's high accuracy, achieving correct atom mapping in 99.4\% of tested reactions, including diverse reaction types such as Diels-Alder reactions, methylene transfers, and epoxidations \cite{schwaller_extraction_2021}. However, the model occasionally demonstrates inaccuracies, particularly regarding atom ordering within rings, azide compounds, and the mapping of oxygen atoms in reductions or Mitsunobu reactions.

\subsection*{Base Case: Entirely Fossil Feedstock}\label{sec:SIbasecase}
\begin{figure}[h]
    \centering
    \includegraphics[width=0.48\textwidth]{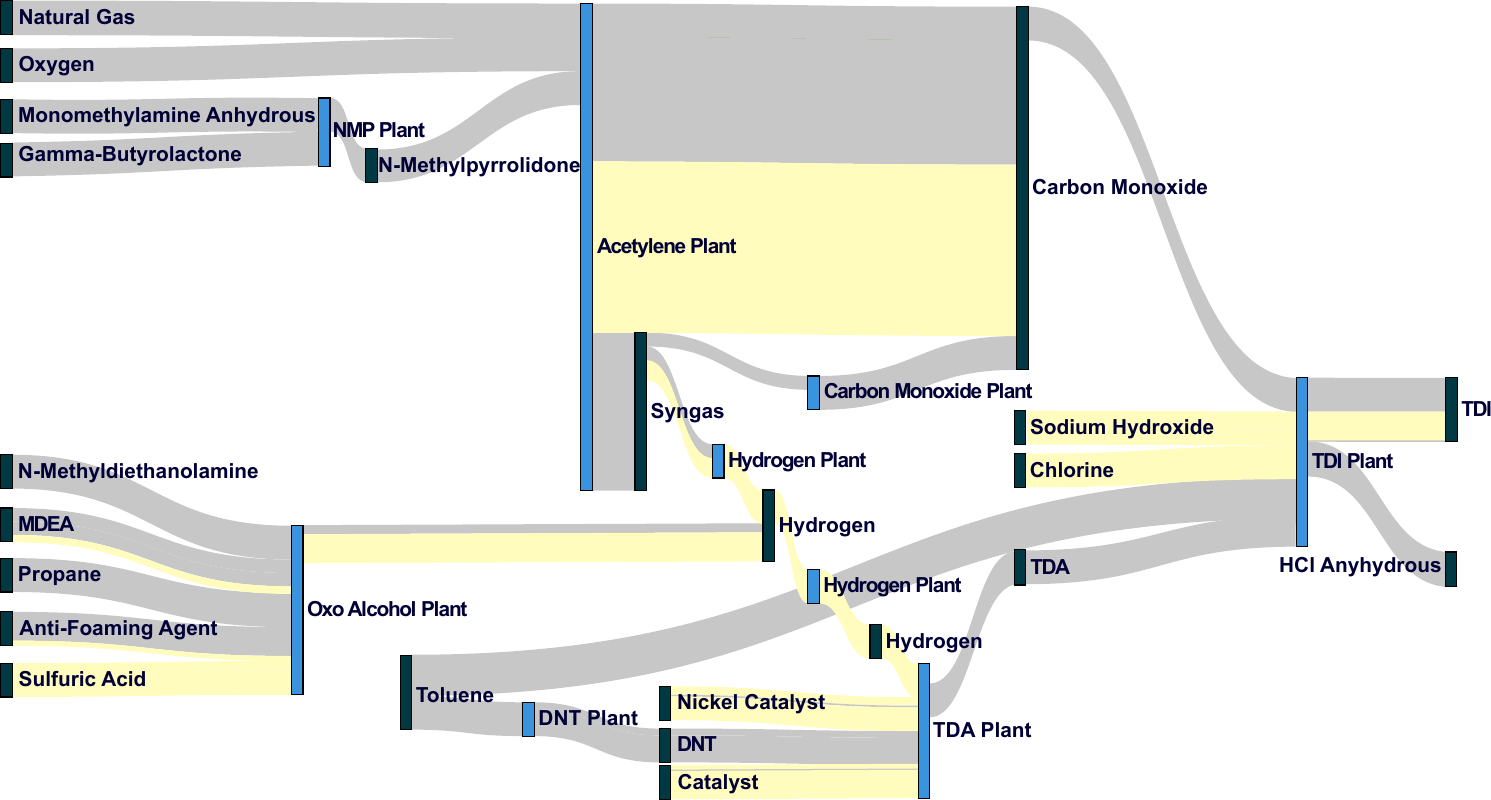}
    \caption{Sankey diagram visualizing results of base case linear program solution. All carbon is fossil-sourced}
    \label{fig:example}
\end{figure}

In the Base Case, all carbon inputs are fossil-derived, giving a product BCC of 0\%. The Sankey diagram confirms that carbon flows track exclusively through gray links; no biogenic carbon enters the chain.

The base case scenario with the TDI value chain serves as the first proof of concept for the framework. While developing the linear program optimization for the final stage of the framework, a step-by-step approach was taken to ensure the validity of the method. The first use of the linear program was at the scale of one node, allowing the results to be verified by hand to ensure that the optimization program and implementation were working as expected. 

After confirming this, the next step was to trial a control study, or base case, in which all carbon is fossil-derived. The purpose of this was to check that the linear program was well-posed and that the results were as expected. This also provided an opportunity to consider how best to represent the results, as the immediate output of the linear program is often unintuitive.
\end{document}